\documentstyle[pra,aps,psfig]{revtex}
\begin{document}          
\draft

%\onecolumn
\twocolumn[\hsize\textwidth\columnwidth\hsize\csname @twocolumnfalse\endcsname
\title{Dark soliton states of Bose-Einstein condensates in anisotropic traps}
\author{D. L. Feder,$^{1,2}$ M. S. Pindzola,$^3$ L. A. Collins,$^4$
B. I. Schneider,$^5$ and C. W. Clark$^{2}$}
\address{$^1$Clarendon Laboratory, University of Oxford, Parks Road, Oxford
OX1 3PU, U.K.}
\address{$^2$Electron and Optical Physics Division, National Institute of
Standards and Technology, Gaithersburg, MD 20899-8410}
\address{$^3$Department of Physics, Auburn University, Auburn, AL 36849}
\address{$^4$Theoretical Division, Mail Stop B212, Los Alamos National
Laboratory, Los Alamos, NM 87545}
\address{$^5$Physics Division, National Science Foundation, Arlington,
Virginia 22230}
\date{\today}
\maketitle

\begin{abstract}

Dark soliton states of Bose-Einstein condensates in harmonic traps are studied
both analytically and computationally by the direct solution of the
Gross-Pitaevskii equation in three dimensions. The ground and self-consistent
excited states are found numerically by relaxation in imaginary time. The
energy of a stationary soliton in a harmonic trap is shown to be independent
of density and geometry for large numbers of atoms. Large amplitude field
modulation at a frequency resonant with the energy of a dark soliton is found
to give rise to a state with multiple vortices. The Bogoliubov excitation
spectrum of the soliton state contains complex frequencies, which disappear
for sufficiently small numbers of atoms or large transverse confinement. The
relationship between these complex modes and the snake instability is
investigated numerically by propagation in real time.

\end{abstract}
]
%\pacs{03.75.Fi, 05.45.Yv, 42.50.-p}

\narrowtext

\section{introduction}
\label{sec:introduction}

Numerous experimental studies have confirmed the general validity of the
time-dependent Gross-Pitaevskii (GP) equation~\cite{a1,a2} used to calculate
the ground state and excitations of various Bose-Einstein condensates
of trapped alkali atoms~\cite{a3,Dalfovo,Ketterle}. To map the spectrum of
collective (or particle-hole) excitations, mean-field linear response
theories based on the Bogoliubov approximation~\cite{a4,a5} or its finite
temperature extensions~\cite{Griffin,Hutchinson} have been
developed and applied to numerous experimental configurations.

The collective excitations are physically distinct from {\it self-consistent}
excited states of the trapped gas. In the latter case, the stationary 
condensate wavefunction itself may contain one or more nodes. Indeed, the
nonlinear GP equation supports many well-known self-consistent excitations,
such as vortex states~\cite{Dodd,Rokhsar,Fetter,Feder1,Castin}, and
configurations with bright and dark
solitons~\cite{Reinhardt,Dum,Zobay,Anglin,Muryshev} for attractive as well as
repulsive Bose gases. These contrast quite strongly with the collective
excitations, which are obtained from the linear response of the condensate to
an external perturbation.

In the case of a fundamental dark (or black) soliton, the condensate density
vanishes along a nodal surface and the soliton velocity is zero. Such a
solution is equivalent to two condensates with a phase difference of $\pi$
between them, separated by a thin impenetrable barrier, and is an idealization
of the nodal structures obtained recently in a two-component
system~\cite{Matthews}. Dark optical solitons in nonlinear dielectric fibers
have been actively studied~\cite{Kivshar} since their
prediction~\cite{Hasegawa} and experimental
observation~\cite{Emplit,Krokel,Weiner}; the recent observation of solitons in
trapped Bose gases~\cite{Ertmer,Denschlag} has provided another striking
manifestation of nonlinear atom optics~\cite{Deng}.

The stability of stationary dark solitons (also known as standing waves or
kinks) in trapped condensates has been the subject of recent
investigations~\cite{Muryshev}. These states are thermodynamically unstable,
since their energies are always higher than the nodeless self-consistent
ground state. In addition, they can be dynamically unstable; in more than
one dimension, an extended dark soliton in an optical fiber will generally
undergo a `snake deformation', where transverse modulations cause the nodal
plane to decay into vortices~\cite{Kivshar}. The Bogoliubov excitation
spectrum for a kink is known to contain modes with imaginary frequencies and
quasiparticle amplitudes localized in the notch~\cite{Muryshev}; however, the
origin of these imaginary eigenvalues and their explicit connection to a
dynamical snake instability remain unclear.

In the present work, the properties and stability of self-consistent excited
states of trapped Bose condensates are explored further. After a brief
description in Sec.~\ref{sec:background} of the formalism and techniques
employed in the numerical calculations, the number-dependence of the energy of
stationary dark solitons is obtained and discussed in Sec.~\ref{sec:stationary}.
In the Thomas-Fermi (TF) limit, corresponding to large condensates, the 
energy difference between the kink and nodeless ground states is found to be
independent of the number of atoms. In order to better understand this result,
the soliton energy is calculated perturbatively around the TF limit using a
boundary-layer approach in Sec~\ref{sec:boundary}. It is shown that the energy
of the soliton state in the TF limit is identical to that of the `anomalous
mode' in the Bogoliubov spectrum. Perturbation theory in the weakly interacting
limit, carried out in Sec.~\ref{sec:weakly}, demonstrates that this result is
particular to large condensates, however. This perturbative approach also
yields significant insight into the criteria for the existence of Bogoliubov
excitations with complex frequencies, discussed in Sec.~\ref{sec:weakly2}. The
relationship between complex modes and dynamical instability of the kink is
explored in Sec.~\ref{sec:snake}. In Sec.~\ref{sec:field}, we explore the
possibility of transferring the condensate into a kink state by a field
excitation. The results are summarized in Sec.~\ref{sec:conclusions}.

\section{theoretical background}
\label{sec:background}

At zero temperature, the dynamics of a single-component condensate are
governed by the three-dimensional (3D) time-dependent GP equation:
\begin{equation}
  i{\partial\psi({\bf{r}},t)\over\partial t}
    = \left( -{1\over 2}\nabla^2 + V_{\rm trap}({\bf{r}}) + V_{\rm H}({\bf{r}},t)
		\right) \psi({\bf{r}},t),
\label{GP}
\end{equation}
where the confining harmonic potential
\begin{equation}
V_{\rm trap}({\bf{r}}) = {1\over 2}(x^2 +\alpha^2 y^2 +\beta^2 z^2)
\end{equation}
is completely anisotropic in general; in recent experiments on
solitons in a Bose condensate~\cite{Denschlag}, the relevant parameters were
$\alpha\equiv\omega_y/\omega_x=\sqrt{2}$ and $\beta\equiv\omega_z/\omega_x=2$.
The Hartree (mean-field) potential is written
\begin{equation}
 V_{\rm H}({\bf{r}},t) = 4\pi\eta_0 |\psi({\bf{r}},t)|^2;
\end{equation}
choosing the condensate wavefunction $\psi({\bf{r}},t)$ to be normalized to
unity yields the strength parameter $\eta_0 = aN_0/d_x$, where $a$ is the
atomic scattering length and $N_0$ is the total number of atoms. We assume a
condensate composed of Na atoms, in which case $a=52a_B\approx 2.75$~nm in Bohr
radii $a_B$~\cite{Eite}. The above three equations are written in reduced
units, where the length scale is $d_x = \sqrt{\hbar/M\omega_x}$, the time scale
is $T = 2\pi/\omega_x$, and the energy is given in units of $\hbar\omega_x$,
where $\omega_x$ is the angular trap frequency in the $x$ direction and $M$ is
the atomic mass of Na.

The ground and self-consistent excited states of Bose-Einstein condensates are
obtained by direct solution of the GP equation in imaginary time ($\tau=it$).
At each imaginary time step, the chemical potential
$\mu\equiv\langle H\rangle/N_0$ (where $H$ is the GP operator on the right
side of Eq.~(\ref{GP})] is readjusted in order to preserve the norm of the
wavefunction. Self-consistent excitations may be found numerically by
relaxation of the GP equation toward equilibrium, subject either to special
initial conditions (spatial variations of phase or amplitude~\cite{Feder1,a13})
or applied constraints (such as orthogonality to the ground state).

The initial wavefunction for the imaginary time propagation is a Gaussian
$\psi({\bf{r}},0)=f({\bf{r}})\exp\{-{\case 1/2}(x^2+\alpha^2y^2+\beta^2z^2)\}$
for small numbers of atoms $N_0\lesssim 10^5$. For larger $N_0$ the kinetic
energy contribution to the total energy becomes negligible, and the initial
state is chosen to be proportional to the Thomas-Fermi (TF) expression
$f({\bf{r}})\sqrt{(\mu_{\rm TF}-V_{\rm trap}) /4\pi\eta_0}\Theta(\mu_{\rm TF}
-V_{\rm trap})$. The TF chemical potential is
$\mu_{\rm TF}={\case 1/2}(15\alpha\beta\eta_0)^{2/5}$ in units of
$\hbar\omega_x$ and $\Theta(x)$ is unity when $x$ is positive and zero
otherwise. The choice of initial state has no influence on the final result,
but can improve the time required for numerical convergence. All stationary
states without circulation can be classified by their reflection symmetry in
the $\hat{x}$, $\hat{y}$, and $\hat{z}$ directions. For convenience,
wavefunctions that are odd under a reflection in one spatial direction
$\alpha=x,y,z$ are labelled $p_{\alpha}$ and are referred to as `p-wave'.
Similarly, the `d-wave' states $d_{\alpha\beta}$, with $\alpha,\beta=x,y,z$,
have odd reflection symmetry in two directions $\alpha$ and $\beta$. In order
to obtain p-wave or d-wave states, one may choose $f({\bf{r}}) \propto x,y,z$
or $f({\bf{r}}) \propto xy,xz,yz$, respectively. Since the nonlinear
Hamiltonian~(\ref{GP}) commutes with the parity operators, however, perhaps
the simplest strategy is to block diagonalize the GP operator according to
parity in each direction, and solve for the lowest energy state in each of the
eight parity manifolds. For higher lying stationary configurations, the
solution must be explicitly orthogonalized to lower-energy states during the
imaginary time propagation.

We propagate the 3D time-dependent GP equation using three distinct
techniques: a variable stepsize Runge-Kutta (RK) method~\cite{NR},
second-order differencing (SOD)~\cite{SOD}, and real-space product formula
(RSPF)~\cite{RSPF}. In the SOD and RK methods, the time propagation results
from one or at most a few matrix-vector multiplies of the Hamiltonian onto a 
previously computed vector. The RSPF employs a split-operator approach which
partitions the action of the kinetic energy matrix into a succession of
$2\times 2$ matrix operations on a known vector. It should be noted that all
of these time propagators are efficiently implemented on distributed-memory
massively parallel computers.

The treatment of the kinetic energy operator forms the main difference in the
implementation of the propagation techniques. The SOD and RSPF methods
discretize the kinetic energy operator using the simplest three-point, central
finite difference (FD) formula for the second derivative operator. In the RK
method, the spatial wavefunction is expanded in a discrete variable
representation (DVR)~\cite{Light,a12} based on Gauss-Hermite quadrature. The
DVR has the advantage, shared also by the FD method, that the matrix elements
of all local operators are diagonal and equal to their value on the spatial
grid. The DVR kinetic energy operator is dense in each dimension compared with
the FD approach; however, it also provides a much more accurate representation
of the derivatives than does the simple FD approximation. All of the methods
scale formally with the number of spatial grid points, although the prefactor
is different for each. Grids of the order of 200$^3$ points are used in the
SOD and RSPF, while the DVR approach employs approximately 100 functions in
each spatial direction.

\section{stationary states}
\label{sec:stationary}

\begin{figure}
\begin{center}
\psfig{figure=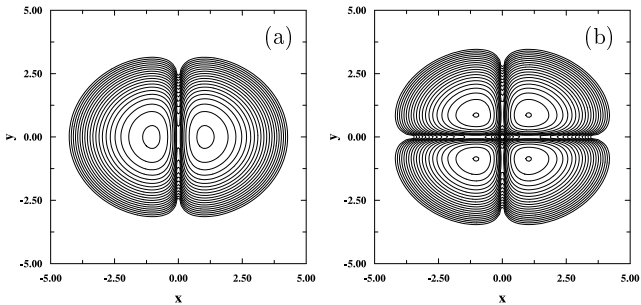,width=\columnwidth,angle=0}
\end{center}
\caption{The probability densities in the $z = 0$ plane for (a) $p_x$ and (b)
$d_{xy}$ dark soliton states with $N = 2^{10}=1024$ are shown as 2D contour
maps, where radial distances are in scaled trap units $d_x$. Trap parameters
are $\omega_x=(2\pi)177$~rad/s, $\alpha=\sqrt{2}$, and $\beta=2$.}
\label{pdwave}
\end{figure}

Using the diffusion form of Eq. (\ref{GP}) for the condensate wavefunction,
the GP equation relaxes to the stationary state of interest. The chemical
potential $\mu$ and free energy per particle
$E\equiv\mu-{\case 1/2}\langle V_{\rm H}\rangle/N_0$ may be obtained as a
function of the number of atoms, and the results are summarized in
Tables~\ref{ptable} and \ref{dtable}. The values of the chemical potential for
the ground state agree to three significant figures with those reported
earlier~\cite{a12}. The probability densities in the $z = 0$ plane for $p_x$
and $d_{xy}$ dark soliton condensates with $N_0 = 2^{10} = 1024$ are shown in
Fig.~\ref{pdwave}. The line nodes, which are clearly visible as depressions in
the condensate density, widen near the surface as the condensate density
decreases and the healing length diverges.

If the GP equation is relaxed subject to orthogonality constraints with
previous solutions, many additional stationary states may be found. The first
s-wave excited state, with $f({\bf{r}}) = 1$ and constrained to be orthogonal
to the ground state, has an ellipsoidal nodal surface centered about zero. The
excited state with $f({\bf{r}}) = 2x^2 -1$ and constrained to be orthogonal to
the ground state, has two nodal surfaces which intersect the $x$ axis.

The numerical calculations indicate that the average energy per particle for a
black soliton is independent of both geometry and particle number in the TF
limit; this is in contrast with the energy per particle of an isolated vortex
in a cylindrical trap, for example, which varies as
$\Delta E_v\sim(5/2R^2)\ln(R/\xi)$ where $R=(15\eta_0)^{1/5}$ and $\xi\sim 1/R$
are respectively the mean TF radius and the healing length in units of
$d_x$~\cite{Fetter}. For large $N_0$, the energy differences (both in the
chemical potential and free energy) between the $p_x$ and ground states
converge to a constant value of approximately $0.7$ in units of
$\hbar\omega_x$. The energy differences of the $p_y$ and $p_z$ states are
simply scaled by $\alpha$ and $\beta$, respectively; i.e.\ the soliton energy
is $\Delta E_s\approx 0.7$ in units of $\hbar\omega_y$ and $\hbar\omega_z$.
Similarly, the energies of the $d_{\alpha\beta}$ states are approximately
$\Delta E_s\approx 0.7(\alpha+\beta)$.

The energy of the black soliton relative to the ground state is close to the
energy $\hbar\omega/\sqrt{2}$ of the `anomalous mode' in the Bogoliubov
spectrum of a one-dimensional p-wave state in the TF limit~\cite{Muryshev}.
This excitation has positive energy but negative norm (or vice versa), and has
been associated with the oscillation of a dark soliton in the trapped
condensate at frequency $\omega/\sqrt{2}$~\cite{Anglin,Denschlag}. In nonlinear
systems, however, there is no direct relationship between the energy
differences among self-consistent states and the collective excitations from
these states; for example, the precession frequency (or anomalous mode) for an
isolated vortex in a cylindrical TF condensate is not $\Delta E_v/\hbar$, but
rather ${\case 3/5}\Delta E_v/\hbar$~\cite{Fetter}. Indeed, as shown below, the
energy of a black soliton displaced from the trap center is always smaller than
$\hbar\omega/\sqrt{2}$.

\section{Perturbative analysis}

\subsection{Thomas-Fermi limit}
\label{sec:boundary}

The soliton energy may be calculated using a boundary layer correction to the
TF wavefunction~\cite{Anglin,Feder3}. The $p_x$ state requires a plane of nodes
at $x_0=0$, but in principle $x_0$ can take any value since an oscillating
dark soliton becomes black at its classical turning
point~\cite{Anglin,Denschlag}. The condensate density will have the TF form
everywhere except in a small region near $x_0$, where the kinetic energy drives
the wavefunction to zero.

In order to define a suitable perturbation parameter, it is convenient to
further rescale the static GP equation~(\ref{GP}), where the left side
becomes $\tilde{\mu}\psi(\vec{r})$, as follows:
$\{x,y,z\}=R\{\tilde{x},\tilde{y}/\alpha,\tilde{z}/\beta\}$,
$\tilde{\mu}=\mu/\hbar\omega_x=R^2/2$, $\tilde{\eta}_0=\eta_0\alpha\beta/R^5$,
and $\psi^2=\tilde{\psi}^2\alpha\beta/R^3$. Now, the normalization of the
wavefunction takes the form
$4\pi\tilde{\eta}_0=\int d^3\tilde{r}\tilde{\psi}^2$. For a condensate in the
TF limit with a soliton normal to the $x$-axis, one may neglect the kinetic
energy contributions except in the $x$-direction. The rescaled GP equation then
effectively has cylindrical symmetry:
\begin{equation}
\left[-{\epsilon\over 2}{\partial^2\over\partial x^2}+{1\over 2}\tilde{r}^2
+\tilde{\psi}^2-{1\over 2}\right]\tilde{\psi}=0,
\label{GPb}
\end{equation}
where $\epsilon=R^{-4}$ is small in the TF limit, and
$\tilde{r}^2=\tilde{x}^2+\tilde{y}^2+\tilde{z}^2=\tilde{x}^2+\tilde{\rho}^2$.

In the outer (slowly-varying) region, one expands $\tilde{\psi}_{\rm out}
=\chi_0+\epsilon\chi_1+\ldots$. The TF result is recovered to zeroth order in
$\epsilon$: $\chi_0=\pm\sqrt{(1-\tilde{r}^2)/2}$. Of course, this solution is
inconsistent with the boundary condition
$\tilde{\psi}(\tilde{x}=\tilde{x}_0)=0$, implying the existence of a boundary
layer near $\tilde{x}\sim\tilde{x}_0$. In this region, the outer solution is
asymptotically $\chi_0\sim\pm\sqrt{(1-\tilde{\rho}^2-\tilde{x}_0^2)/2}$.

For the inner region it is preferable to define
$\tilde{x}\equiv\tilde{x}_0+\delta X$, where the boundary-layer thickness is
$\delta\ll 1$ and $|X|\gg 0$ are the regions where the inner and outer
solutions must match. Since the asymptotic behavior of the outer solution is
known, the inner wavefunction may be expanded as
$\tilde{\psi}_{\rm in}=\pm\sqrt{(1-\tilde{\rho}^2-\tilde{x}_0^2)/2}
\left[\Phi_0+\delta\Phi_1+\ldots\right]$. To lowest order in $\delta$,
Eq.~(\ref{GPb}) becomes:
\begin{equation}
\left[{\epsilon\over\delta^2}{\partial^2\over\partial X^2}
+\left(1-\tilde{\rho}^2-\tilde{x}_0^2\right)\left(1-\Phi_0^2\right)\right]
\Phi_0=0.
\label{GPin}
\end{equation}
The `distinguished limit' giving a nontrivial solution corresponds to
$\delta=\sqrt{\epsilon}=R^{-2}$. When $|X|\gg 0$, $\Phi_0\sim 1$ giving a
perfect asymptotic match. With the substitution
$X=Z/\sqrt{{\case1/2}(1-\tilde{\rho}^2-\tilde{x}_0^2)}$, Eq.~(\ref{GPin})
becomes the well-known equation for a dark soliton in the continuum
$-{\case1/2}{\case{d^2}/{dZ^2}}\Phi_0+\Phi_0^3-\Phi_0=0$, yielding the exact
solution for the inner wavefunction
$\Phi(X)=\tanh\left[X\sqrt{{\case1/2}(1-\tilde{\rho}^2-\tilde{x}_0^2)}\right]$.

The uniform solution for the wavefunction over all space may be written as
$\tilde{\psi}_{\rm unif}=\tilde{\psi}_{\rm out}+\tilde{\psi}_{\rm in}
-\tilde{\psi}_{\rm over}$, where $\tilde{\psi}_{\rm over}$ is the solution in
the overlap region $|X|\gg 0$:
\begin{eqnarray}
&&\tilde{\psi}_{\rm unif}(\tilde{\rho},\tilde{x}_>^<)=\mp\sqrt{1-\tilde{\rho}^2
-\tilde{x}^2\over 2}+\sqrt{1-\tilde{\rho}^2-\tilde{x}_0^2\over 2}\nonumber \\
&&\qquad\times\left\{\tanh\left[\sqrt{1-\tilde{\rho}^2-\tilde{x}_0^2\over 2}R^2
(\tilde{x}-\tilde{x}_0)\right]\pm 1\right\},
\label{bl}
\end{eqnarray}
where the $\tilde{x}_<$ and $\tilde{x}_>$ correspond to $\tilde{x}<\tilde{x}_0$
and $\tilde{x}>\tilde{x}_0$, respectively. In principle, the chemical potential
may now be found directly from the normalization condition
$4\pi\tilde{\eta}_0=\int d^3\tilde{r}\tilde{\psi}_{\rm unif}^2$. In practice,
however, the large number of cross terms resulting from squaring Eq.~(\ref{bl})
makes this unnecessarily complicated. Rather, the integral over $\tilde{x}$ is
split into three regions:
(I) $-\sqrt{1-\tilde{\rho}^2}\leq\tilde{x}\leq\tilde{x}_a$
($\tilde{x}_a<\tilde{x}_0$), (II) $\tilde{x}_a\leq\tilde{x}\leq\tilde{x}_b$
($\tilde{x}_b>\tilde{x}_0$) or $X_a\leq X\leq X_b$ ($X_a\to -\infty$ and
$X_b\to\infty$), and (III) $\tilde{x}_b\leq\tilde{x}\leq
\sqrt{1-\tilde{\rho}^2}$. Since the inner and outer solutions are
asymptotically equal in the overlap regions, the result cannot depend on the
particular choices of $\tilde{x}_a$ and $\tilde{x}_a$. One readily obtains:
\begin{equation}
\tilde{\eta}_0={\alpha\beta\eta_0\over R^5}={1\over 15}-{\sqrt{2}\over 6R^2}
\left(1-\tilde{x}_0^2\right)^{3/2},
\end{equation}
which may be inverted to yield the chemical potential for the soliton state
\begin{eqnarray}
\tilde{\mu}_s&=&{\mu_s\over\hbar\omega_x}={1\over 2}\left(15\alpha\beta\eta_0
\right)^{2/5}+{1\over\sqrt{2}}\left(1-\tilde{x}_0^2\right)^{3/2}\nonumber \\
&=&\tilde{\mu}_{\rm TF}+{1\over\sqrt{2}}\left(1-\tilde{x}_0^2\right)^{3/2}.
\label{blmu}
\end{eqnarray}
The correction to the chemical potential in the TF limit is $1/\sqrt{2}$ when
the soliton is at the origin $\tilde{x}_0=0$, corresponding to the p-wave
state, and is zero when the soliton is at the surface of the cloud
$\tilde{x}_0=1$. Since $\mu=\partial E/\partial N$, the energy per particle
for the soliton is
$${E_s\over N_0}={5\over 7}\mu_{\rm TF}+{1\over\sqrt{2}}
\left(1-\tilde{x}_0^2\right)^{3/2}\hbar\omega_x.$$

The corrections for $\tilde{x}_0=0$ agree with the numerical values, and are
independent of both geometry and number at this level of approximation.
Physically, the dark soliton has a constant energy as the number of atoms
increases, because its area increases as $R^2$ while its width diminishes as
$\xi^2\sim 1/R^2$~\cite{Fetter}. A similar invariant is used as a measure of
soliton stability in optical fibers~\cite{Tomlinson}. As the transverse
soliton confinement becomes more appreciable (i.e.\ by increasing the
frequencies $\omega_y$ and $\omega_z$), however, the kinetic energy in this
direction will grow, and the low-order boundary-layer result will lose its
validity.

\begin{figure}
\begin{center}
\psfig{figure=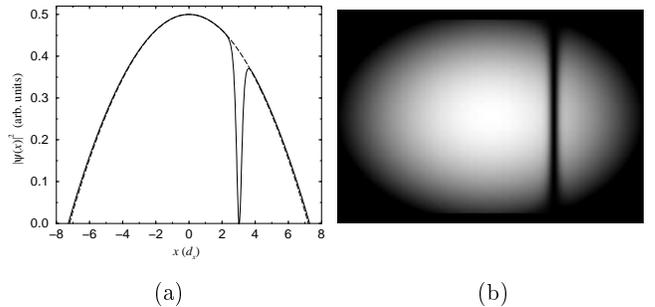,width=\columnwidth,angle=0}
\end{center}
\caption{The boundary-layer approximation~(\ref{bl}) for a condensate
containing a soliton at $x_0=3d_x$ is shown for $N_0=2^{18}=262,144$ atoms,
$\omega_x=(2\pi)177$~rad/s, $\alpha=\sqrt{2}$, and $\beta=2$. In (a), the view
is along the $x$-axis; the solid and dashed lines correspond to the soliton and
TF ground states, respectively. In (b), the soliton state is depicted as a
density plot in the $xy$-plane.}
\label{blfig}
\end{figure}

Two views of a black soliton state defined by Eq.~(\ref{bl}) are shown in
Fig.~\ref{blfig}. In the first, the density with a soliton displaced from the
origin, $x_0=3d_x$, is shown along the $x$-axis, and is compared with the TF
ground state solution. The condensate containing a notch soliton bulges
slightly overall in order to conserve the total number of atoms; the radii for
the TF and soliton states are $d_x\sqrt{2\tilde{\mu}_{\rm TF}}$ and
$d_x\sqrt{2\tilde{\mu}_s}$, respectively. In the second view, the soliton
state is shown as a density plot in the $xy$-plane. The boundary layer theory
to lowest order captures the divergence of the healing length in the vicinity
of the cloud surface. In actuality, a displaced soliton would most likely be
curved in order for the nodal plane to intersect the cloud boundary at a
surface normal; such curvature is found for travelling dark
solitons~\cite{Denschlag} and for displaced vortices in rotating trapped
condensates~\cite{Feder1,Castin}.

\subsection{Low density limit: Energy differences and anomalous mode}
\label{sec:weakly}

The boundary layer analysis indicates that the energy of the self-consistent
$p_x$ state in the TF limit is exactly equal to the frequency of the anomalous
mode in the Bogoliubov spectrum. In order to determine whether this holds for
all densities, it is useful to consider the opposite limit of small
condensates where analytical results can be easily obtained.

The perturbation expansion for the time-independent GP equation~(\ref{GP})
requires expanding the condensate wavefunction $\psi=\psi_0+\lambda\psi_1$ and
chemical potential $\tilde{\mu}=\tilde{\mu}_0+\lambda\tilde{\mu}_1$ in powers
of $\lambda\equiv 4\pi\eta_0$. The normalized unperturbed $p_x$ state is
\begin{equation}
\psi_0=\left({4\alpha\beta\over\pi^3}\right)^{1/4}x
e^{-\left(x^2+\alpha y^2+\beta z^2\right)/2}
\label{pxcond}
\end{equation}
and $\tilde{\mu}_0={\case1/2}\left(3+\alpha+\beta\right)$. Making use of
the readily derived expression for a purely real condensate wavefunction
\begin{equation}
\tilde{\mu}_1=\int d^3r\psi_0^4={3\over 8\pi}\sqrt{\alpha\beta\over 2\pi},
\label{mupert}
\end{equation}
one immediately obtains the first order correction to the chemical potential
\begin{equation}
\mu\approx\left(\tilde{\mu}_0+{3\over 2}\sqrt{\alpha\beta\over 2\pi}\eta_0
\right)\hbar\omega_x.
\end{equation}

The low-lying excitations $\varepsilon$ may be obtained using the Bogoliubov
equations
\begin{eqnarray}
\left(-{\case1/2}\nabla^2+V_{\rm trap}+2V_{\rm H}-\tilde{\mu}\right)u
-V_{\rm H}v&=&\tilde{\varepsilon}u
\nonumber \\
\left(-{\case1/2}\nabla^2+V_{\rm trap}+2V_{\rm H}-\tilde{\mu}\right)v
-V_{\rm H}u&=&-\tilde{\varepsilon}v,
\label{Bogs}
\end{eqnarray}
where $u=u({\bf r})$ and $v=v({\bf r})$. These equations may be written in the
more convenient form
$\left(H_0+\lambda H_1\right)\psi=\tilde{\varepsilon}\Psi$, where
\begin{eqnarray}
H_0&=&\pmatrix{-{\case1/2}\nabla^2+V_{\rm trap}-\tilde{\mu}_0 & 0 \cr
0 & {\case1/2}\nabla^2-V_{\rm trap}+\tilde{\mu}_0\cr},\nonumber  \\
H_1&=&\pmatrix{2\psi_0^2-\tilde{\mu}_1 & \psi_0^2 \cr
-\psi_0^2 & \tilde{\mu}_1-2\psi_0^2 \cr},\qquad
\Psi=\pmatrix{u \cr v}.
\end{eqnarray}
For arbitrary trap anisotropy, there are always two degenerate modes with
energy $\tilde{\varepsilon}=\varepsilon/\hbar\omega_x=1$ in the unperturbed
$p_x$ state; these are the dipole and anomalous modes with positive and
negative unit norms, respectively:
\begin{eqnarray}
\Psi_1&=&\pmatrix{1 \cr 0}\left({\alpha\beta\over 4\pi^3}\right)^{1/4}
\left(2x^2-1\right)e^{-\left(x^2+\alpha y^2+\beta z^2\right)/2},
\label{dipole} \\
\Psi_2&=&\pmatrix{0 \cr 1}\left({\alpha\beta\over\pi^3}\right)^{1/4}
e^{-\left(x^2+\alpha y^2+\beta z^2\right)/2}.
\label{anomalous}
\end{eqnarray}
Note that the `ground state' is the p-wave condensate wavefunction given in
Eq.~(\ref{pxcond}). The degeneracy between these states is lifted by the
perturbing Hamiltonian $H_1$, and the first-order corrections to the
eigenvalues follow directly from the diagonalization of the resulting
nonsymmetric $g\times g$ matrix
\begin{equation}
\langle\Psi_i|H_1|\Psi_j\rangle,\quad i,j=1,2,\ldots,g,
\label{degen}
\end{equation}
where $g$ is the degeneracy. Direct application of Eq.~(\ref{degen}) with
$g=2$ yields no finite-number correction for the dipole mode, as expected,
while the energy of the anomalous mode becomes
\begin{equation}
\varepsilon_a\approx\left(1-{1\over 4}\sqrt{\alpha\beta\over 2\pi}\eta_0\right)
\hbar\omega_x.
\end{equation}
Evidently, the value of the anomalous mode in the $p_x$ state is not generally
equal to the difference~(\ref{mupert}) in the chemical potentials between the
p-wave and s-wave states
\begin{equation}
\Delta\mu=\left(1+{3\over 2}\sqrt{\alpha\beta\over 2\pi}\eta_0\right)
\hbar\omega_x.
\end{equation}
Indeed, the perturbative corrections do not even have the same sign.

\subsection{Low density limit: Complex modes}
\label{sec:weakly2}

\begin{figure}
\begin{center}
\psfig{figure=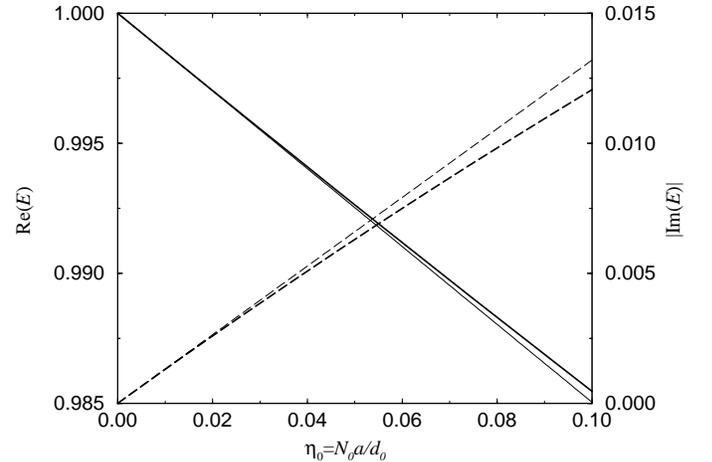,width=\columnwidth,angle=270}
\end{center}
\caption{The real (solid lines) and imaginary (dashed lines) part of the 
complex excitation frequencies for $m=0$ are shown as a function of
$\eta_0=N_0a/d_0$ for a spherical trap. Light and dark lines correspond to 
analytical~(\ref{perturbe}) and numerical calculations, respectively.}
\label{perturb}
\end{figure}

It is instructive to consider the special case where all of the trapping
frequencies are equal, $\alpha=\beta=1$. In this geometry, according to
Ref.~\onlinecite{Muryshev}, an infinitesimal condensate number gives rise to
pure imaginary frequencies. Assuming a $p_x$ state, one may define
$\rho^2\equiv y^2+z^2$ and block diagonalize the Hamiltonian into states of
definite angular momentum $L_x=m\hbar$. In the $m=0$ manifold, in addition to
the dipole~(\ref{dipole}) and anomalous~(\ref{anomalous}) excitations, there
is a third mode with $\tilde{\varepsilon}=1$:
\begin{equation}
\Psi_3=\pmatrix{1 \cr 0}{1\over \pi^{3/4}}\left(1-\rho^2\right)
e^{-\left(x^2+\rho^2\right)/2}.
\label{rhomode}
\end{equation}
Diagonalizing the resulting $3\times 3$ matrix~(\ref{degen}), one again
obtains no correction for the dipole mode. The remaining degenerate modes,
however, split into complex conjugate pairs with energies
\begin{equation}
\varepsilon=\left[1-{1\over 8\sqrt{2\pi}}\left(3\pm i{\sqrt{7}}\right)
\eta_0\right]\hbar\omega_x.
\label{perturbe}
\end{equation}
These results are compared with numerical calculations for a spherical trap in
Fig.~\ref{perturb}. The numerics are extremely close to the
expression~(\ref{perturbe}) for small $\eta_0$, but show deviations by
$\eta_0\sim 0.1$.

The corresponding complex eigenvectors both have zero norm
$\int d^3r\left(|u_i|^2-|v_i|^2\right)\equiv\int d^3r\Psi_i^*\sigma_3\Psi_i=0$,
and satisfy the boundary conditions $u,v\to 0$ as ${\bf r}\to\infty$:
\begin{eqnarray}
\Psi_2'&=&\Psi_1-\sqrt{2}\Psi_2-{\sqrt{2}\over 4}\left(1-i\sqrt{7}\right)\Psi_3;
\\
\Psi_3'&=&\Psi_1-\sqrt{2}\Psi_2-{\sqrt{2}\over 4}\left(1+i\sqrt{7}\right)\Psi_3.
\end{eqnarray}
The condensate coupling between the axial and radial modes gives rise to modes
with frequencies that are {\em complex}, rather than purely imaginary as
assumed in~\onlinecite{Muryshev}. It is important to note that the existence
of complex Bogoliubov excitations does not violate the general condition on
the quasiparticle amplitudes~\cite{a5}:
\begin{equation}
\left(\varepsilon_i-\varepsilon_j^*\right)\int d^3r\left(u_j^*u_i-v_j^*v_i
\right)=0.
\label{proof}
\end{equation}
If $i=j$ and $\varepsilon_i$ is complex, the corresponding particle-hole
eigenfunction must have zero norm.

For any given number of condensate atoms $N_0$ in a cylindrically symmetric
trap, there is a critical anisotropy $\omega_{\rho}/\omega_x=\alpha$ such that
all the Bogoliubov excitations of the p-wave state become purely
real~\cite{Muryshev}. Indeed, in the limit $N_0\to 0$ considered here, any
$\alpha>1$ is sufficient to ensure the disappearance of the complex modes
because the degeneracy between the anomalous and transverse modes is broken by
the anisotropy.

For $\alpha<1$, complex modes can arise in many $m$ states. In this regime,
there exist numerous additional unperturbed anomalous modes with energy
$\tilde{\varepsilon}_{nm}=1-\alpha(2n+m)\geq 0$ but negative norm
(since $u=0$) that are degenerate with eigenstates having energy
$\tilde{\varepsilon}_{n'm'}=\alpha(2n'+m')-1\equiv\tilde{\varepsilon}_{nm}$
and positive norm ($v=0$). It is interesting to determine if the nonlinear
coupling among these degenerate modes gives rise to complex excitations even
in the limit of vanishing transverse confinement, $\alpha\to 0$.
Consider the $m=0$ manifold and $\alpha=1/q$ with even integer $q\to\infty$.
Degeneracies occur when the axial quantum numbers for both the $u$ and $v$ are
zero and the radial numbers $n_u={\case{q}/2}+p$ and $n_v={\case{q}/2}-p$,
respectively, where $0\leq p\leq{\case{q}/2}$. When $p={\case{q}/2}$, the
unperturbed energy is $\tilde{\varepsilon}_0=1$, but since the terms in
Eq.~(\ref{degen}) involving $\Psi_3\propto L_q^{(0)}/q!$ will be smaller than
the other terms by a factor $1/q!$, this complex mode vanishes in the limit
$q\to\infty$. In the opposite limit $p=0$, the unperturbed energy is
$\tilde{\varepsilon}_0=0$, and the quasiparticle amplitudes become
\begin{eqnarray}
u=v&=&{L_{q/2}^{(0)}(\rho/\sqrt{q})\over\pi^{3/4}\sqrt{q}(q/2)!}
e^{-(\rho^2/q+x^2)/2}\nonumber \\
&\sim&J_0(\sqrt{2}\rho)e^{-x^2/2},\quad q\to\infty.
\end{eqnarray}
Note that this $u=v$ solution has even $x$-symmetry and therefore does not
correspond to the Goldstone mode. The condensate wavefunction $\psi$ in
Eq.~(\ref{pxcond}) becomes independent of $\rho$, and the $2\times 2$
matrix~(\ref{degen}) immediately yields the imaginary eigenvalues
\begin{equation}
\varepsilon=\pm i\sqrt{3\pi\over 2}n_0ad_x^2\hbar\omega_x,
\label{immodes}
\end{equation}
where $n_0=N_0/{\cal V}$ is the condensate density in the system volume
${\cal V}$. The same solution is found for all values of $m$ in this limit.
Thus, even in the absence of transverse confinement, the excitation spectrum
of the $p_x$ state contains complex modes, in agreement with the results of
Ref.~\onlinecite{Muryshev}.

Additional insight into the limit of transverse deconfinement may be gained by
assuming translational invariance in $\hat{x}$ and $\hat{y}$ at the outset.
With the axial quantum numbers zero, the unperturbed energies for the $u$ and
$v$ become
$\tilde{\varepsilon}_0^u={\case1/2}\left({k_y^u}^2+{k_z^u}^2\right)-1$ and
$\tilde{\varepsilon}_0^v=1-{\case1/2}\left({k_y^v}^2+{k_z^v}^2\right)$.
When these are degenerate, the off-diagonal couplings are non-zero only if
${k_y^u}={k_y^v}=k_y$ and ${k_z^u}={k_z^v}=k_z$, enforcing the condition
$\tilde{\varepsilon}_0^u=\tilde{\varepsilon}_0^v=0$ and $k_y^2+k_z^2=k^2=2$.
Note that the unperturbed energy $\varepsilon_0=0$ and relevant wavevector
$|k|=\sqrt{2}$ are the same as in the infinitely weak trap limit considered
above. The perturbing matrix~(\ref{degen}) consists of an infinite number of
identical $2\times 2$ submatrices for a given value of $k_x$ and
$k_y=\sqrt{2-k_x^2}$. Each submatrix corresponds to a different value of $m$
in the cylindrical case considered above, and yields the imaginary modes
\begin{equation}
\varepsilon_k=\pm i\sqrt{3\pi\over 2}n_0ad_x^2{\hbar^2k^2\over 2M}
=\pm i\sqrt{3\pi\over 2}n_0ad_x^2\hbar\omega_x,
\end{equation}
in agreement with the result~(\ref{immodes}). 

Summarizing the results of this section, in the limit of weak particle
interactions or low condensate densities, we have found that the anomalous
mode frequency does not correspond to the frequency difference between the
excited and ground state energies. In addition, solitons are unstable in a
sufficiently loose trap. For solitons oriented in the radial direction of a
cylindrically symmetric trap, imaginary eigenvalues appear in the Bogoliubov
excitation spectrum when the axial frequency begins to exceed the radial
frequency. These modes persist in the limit of vanishing radial confinement.

\section{Complex excitations and snake instability}
\label{sec:snake}

The explicit connection between the existence of excitations with complex
frequencies and the dynamic instability of the p-wave state remains unclear.
This `snake instability' is well known in the nonlinear optical
community~\cite{Kivshar}, and is associated with the undulation of the nodal
plane in the radial direction. It has been conjectured that the complex modes
are responsible for the snake instability~\cite{Muryshev}; however, these
modes have zero norm by definition (\ref{proof}), so it is not clear how they
can become occupied. It is important, therefore, to determine if the wavelength
of the undulation matches the spatial dependence of complex modes in the
Bogoliubov spectrum, and if the snake instability disappears when the
excitation frequencies become purely real.

\begin{figure}
\begin{center}
\psfig{figure=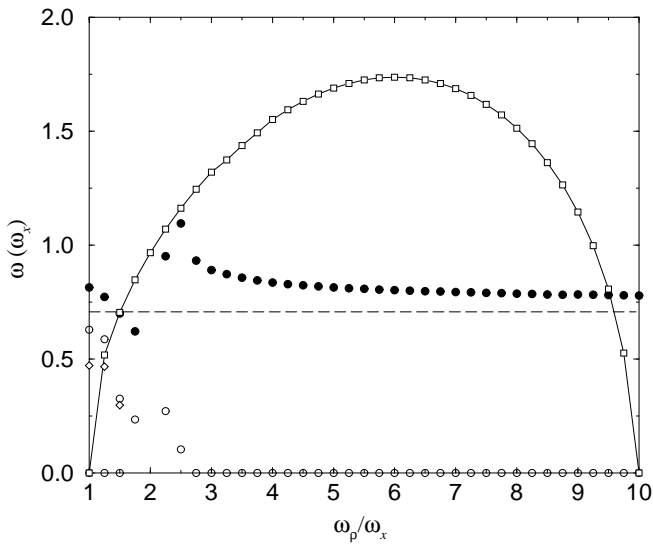,width=\columnwidth,angle=270}
\end{center}
\caption{The real (filled symbols) and imaginary (open symbols) parts of the
low-lying complex excitation frequencies are given as a function of the trap
anisotropy $\alpha\equiv\omega_{\rho}/\omega_x$ for a cylindrically symmetric
trap, where the condensate is in a $p_x$ state with $10^4$ atoms and
$\omega_x=2\pi\cdot 50$~rad/s. Excitations with $m=0$, 1, and 2 are
represented by circles, squares, and diamonds, respectively; the dashed line
denotes the TF estimate of the anomalous mode.}
\label{excfig}
\end{figure}

The low-lying complex modes for a stationary p-wave condensate are determined
numerically using the Bogoliubov equations~(\ref{Bogs}). For ease of
computation, the trap is assumed to be cylindrically symmetric,
$\rho^2=y^2+z^2$, with the condensate wavefunction odd under reflection in
axial direction $\hat{x}$. In Fig.~\ref{excfig} are shown the complex modes in
the excitation spectrum as a function of radial confinement, for a trap with
axial frequency $\omega_x=2\pi\cdot 50$~rad/s containing a condensate with
$10^4$ atoms. All complex modes have even axial parity. For relatively weak
trap anisotropy $\omega_{\rho}/\omega_x\equiv\alpha\lesssim 3$, there are
several complex modes with angular momentum projections $m=0,1,2$. As $\alpha$
increases, these modes disappear in turn until only one purely imaginary mode
with $m=1$ remains. Its magnitude reaches a maximum at $\alpha\approx 6$, and
vanishes at $\alpha\approx 10$. A similar removal of the imaginary modes can
be effected by decreasing the particle number at fixed geometry.

For large anisotropy $\alpha\gg 1$, the stability criterion is expected to be
approximately $\alpha_c\geq\tilde{\mu}/2.4$~\cite{Muryshev}, where
$\tilde{\mu}=\mu/\hbar\omega_x$. In the TF approximation, $\alpha_c\geq 7.6$
for the geometry considered above. The larger value required here is due to
deviations from the TF limit; the radial wavefunction approaches a Gaussian
when the transverse confinement is strong. When $\alpha=10$, the TF chemical
potential is $\mu_{\rm TF}=23.88\hbar\omega_x$ while the actual value is
determined to be $\mu\approx 26.89\hbar\omega_x$.

The dynamic stability of dark soliton excited states is investigated by
propagation of the GP equation in {\em real} time for an extended period. In
the absence of an applied perturbation, the self-consistent states should
remain absolutely stationary. In practice, however, numerical noise inherent
in the propagation algorithm is magnified by the nonlinearity. Although the
norm and chemical potential are conserved to one part in $10^{12}$ and
$10^{15}$ respectively during the numerical propagation, the kink state
eventually decays. In order to make contact with the complex excitation
frequencies in Fig.~\ref{excfig}, we considered the cases $\alpha=1$, 6, and
10. For the first two cases, the modes with the largest imaginary component
are $\varepsilon=0.63i\hbar\omega_x$ and $1.74i\hbar\omega_x$; the p-wave
states were found to decay in real time in approximately 160~ms and 40~ms,
respectively. The third case with $\alpha=10$ remained stable for the longest
propagation time considered, 200~ms. Thus, the lifetime of the p-wave state
appears qualitatively to scale with the inverse of the largest imaginary mode
in the excitation spectrum.

\begin{figure}
\begin{center}
\psfig{figure=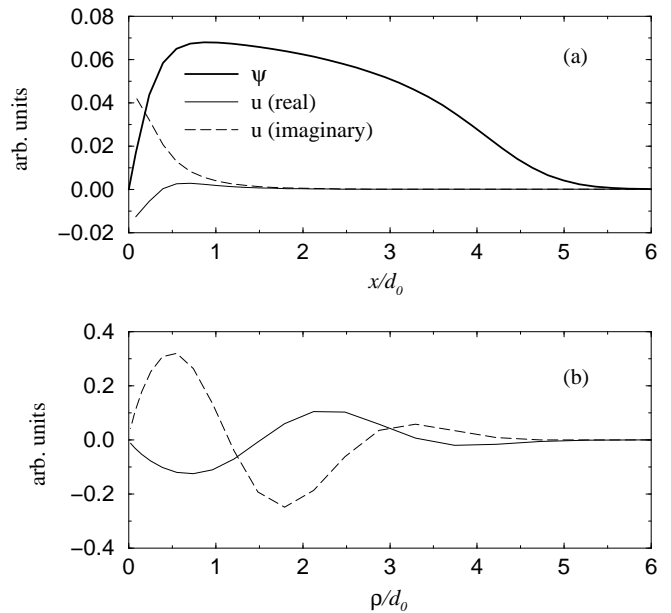,width=\columnwidth,angle=270}
\end{center}
\caption{The spatial variation of the mode in the Bogoliubov spectrum with the
largest imaginary component $\epsilon=2.16i\hbar\omega_x$ is shown for a
$p_x$-wave condensate containing $10^5$ atoms in a spherical trap
$\omega_{\rho}/\omega_x=1$ with $\omega_x=2\pi\cdot 50$~rad/s. The axial
($\rho=0$, $x>0$) and radial ($x=0$) dependences of the complex $u$ amplitude
are shown in (a) and (b), respectively (note that $\psi$ and $u$ are odd and
even in $x$, respectively, and that $|u|=|v|$ for this pure imaginary mode).
The axial dependence of the condensate wavefunction is shown for comparison in
(a).}
\label{imaguv}
\end{figure}

\begin{figure}
\begin{center}
\psfig{figure=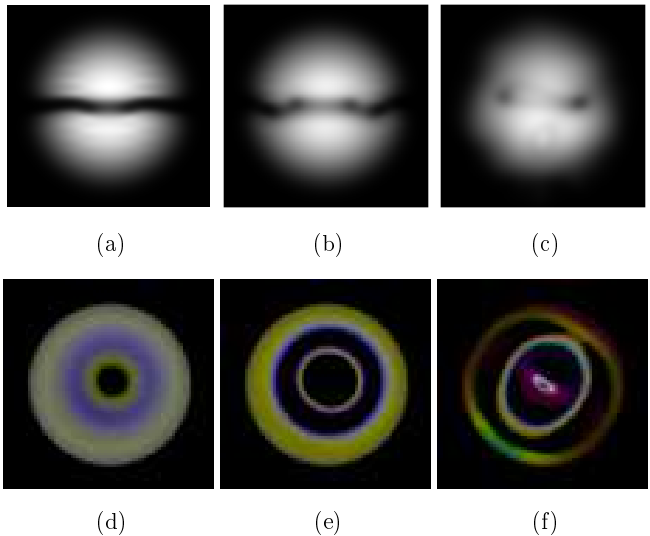,width=\columnwidth,angle=0}
\end{center}
\caption{Snapshots of the snake instability are shown for a $p_y$-wave
condensate containing $10^5$ atoms in a spherical trap with
$\omega=2\pi\cdot 50$~rad/s. Times after the initial formation of the soliton
state are 47~ms, 50~ms, and 77~ms for (a)-(c) and (d)-(f). In (a)-(c), the
brightness is proportional to the condensate density, and the images correspond
to densities integrated down the line of sight. In (d)-(f), the brightness is
{\it inversely} proportional to the condensate density, and regions outside the
TF sphere are rendered transparent in order to visualize nodes in the
condensate interior; the color corresponds to the phase: $\phi=0$ through
$2\pi$ is represented by the sequence red-green-blue-red. The view is
perpendicular to nodal plane; prior to the snake instability the black soliton
would appear as a featureless disk.}
\label{snakesphere}
\end{figure}

As illustrated in Figs.~\ref{imaguv} and \ref{snakesphere}, there is a close
similarity between the spatial variation of the eigenmode with the largest
imaginary component and that of the soliton nodal plane during the initial
decay. For a p-wave state in a sperical trap, such as the one considered above
but with $10^5$ atoms, the relevant excitation is purely imaginary with an
energy $\epsilon=2.16i\hbar\omega_x$, and $|u|=|v|$ (as is the case for all pure
imaginary modes). Fig.~\ref{imaguv} shows the corresponding radial and axial
dependences of the complex $u$ Bogoliubov amplitude. The quasiparticle
amplitudes are highly localized axially, but oscillate radially within the
soliton nodal plane. It is interesting to note that the imaginary components
of the excitation energies tend to {\it decrease} as the number of radial
nodes increases; this behavior is due to the effective negative kinetic energy
of the dark soliton~\cite{Anglin}, and is reminiscent of internal waves at the
interface between layers in stratified fluid mixtures~\cite{Lighthill}.

In Fig.~\ref{snakesphere}, snaphots of the snake instability are shown for a
$p_y$-wave state in a spherical trap [for convenience, the axis perpendicular
to the nodal line is taken to be along $\hat{y}$, the vertical direction in
Fig.~\ref{snakesphere}(a)-(c)]. The GP equation is solved on a Cartesian mesh
with no parity restrictions. After approximately 40~ms of real time
propagation, the black soliton begins to undulate. The spatial variations are
symmetric about the $y$-axis, originating near the center and propagating
outwards. The overall shape follows closely that of the largest $u$ amplitude
shown in Fig.~\ref{imaguv}: the two radial nodes of this imaginary excitation
correspond to stationary points in the soliton bending. At an intermediate
time, Fig.~\ref{snakesphere}(e), the soliton has decayed into two concentric
vortex rings whose cores are located at these nodes. The outer vortex ring
decays rapidly (by 75~ms) to the condensate surface, where it shrinks
considerably, as shown in Fig.~\ref{snakesphere}(f); the inner ring was found
to remain stable for much longer times. The vortex rings are barely visible in
the integrated densities shown in Fig.~\ref{snakesphere}(a)-(c), so we expect
the experimental observation of these intriguing features using standard
absorption imaging to be a challenge. The decay of the soliton into
vortex rings produces a large number of density oscillations, which are
required in order to conserve the total energy of the system and which are
undamped in the present formalism. At long times, these oscillations are most
evident at the condensate surface, giving rise to the bright halo in
Fig.~\ref{snakesphere}(f).

\begin{figure}
\begin{center}
\psfig{figure=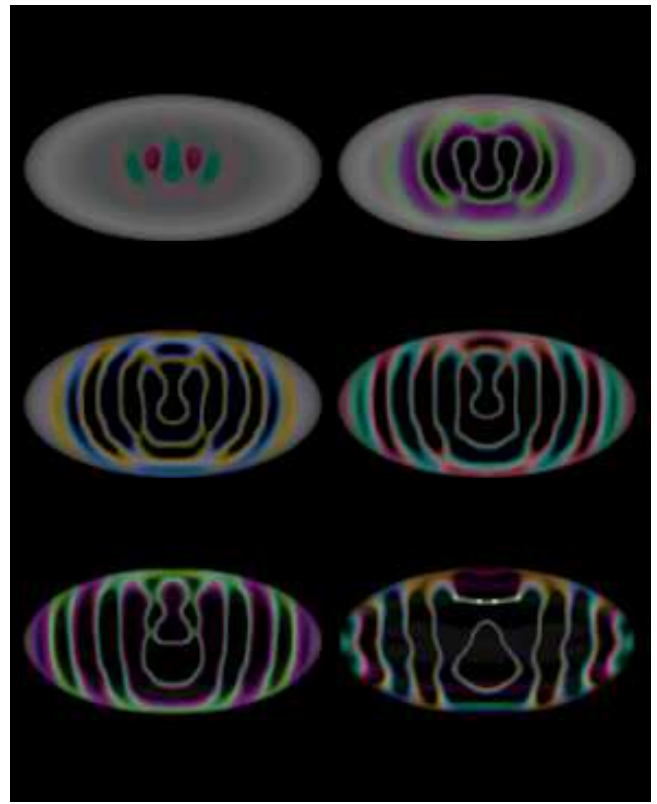,width=\columnwidth,angle=0}
\end{center}
\caption{The breakup of a $p_y$ state is shown as a function of time for
$N_0=10^6$ atoms, $\omega_x=(2\pi)14$~rad/s, $\alpha=\sqrt{2}$, and $\beta=2$.
From the top left to the bottom right in raster order are shown times $t=15$~ms
through $20$~ms in 1~ms increments after the initial state is formed. The view
is along $\hat{y}$, and the Hamiltonian was constrained to even parity along
$\hat{x}$ and $\hat{z}$ for ease of computation. The rendering is identical to
that of Figs.~\ref{snakesphere}(d)-(f). The filamentation is almost entirely
constrained to the original nodal $xz$-plane.}
\label{snake}
\end{figure}

It is interesting to investigate the snake instability for parameters relevant
to recent experiments on dark solitons in trapped condensates~\cite{Denschlag}.
Fig.~\ref{snake} shows the breakup of a black soliton in a completely
anisotropic trap containing one million Na atoms. The undulations are
already pronounced by 12~ms, and radiate outwards from the soliton center as
found above. Unlike the spherically symmetric condensate shown in
Fig.~\ref{snakesphere}, however, the soliton does not decay into concentric
vortex rings far from its center, but rather into a series of approximately
evenly spaced curved vortex lines in the direction of weakest confinement
(taken to be $\hat{x}$). The results imply that the largest imaginary mode in
the Bogoliubov excitation spectrum has an energy of $\sim 6\hbar\omega_x$ and
has 14 nodes along $\hat{x}$. At longer times, the innermost two vortex rings
make contact with one another, and subsequently detach into a vortex line and
a ring. The simulations indicate a rich dynamics among quantized vortices in
these systems that are only beginning to be explored~\cite{Davies}.

\section{Field excitation}
\label{sec:field}

Several techniques have been proposed for the experimental production of 
dark solitons in trapped condensates, including adiabatic Raman transitions
to the p-wave state~\cite{Dum}, preparation of the condensate in a
superposition of two internal states~\cite{Zobay}, collisions between two
separate condensates~\cite{Scott}, and phase imprinting~\cite{Dobrek}; the
last approach was recently implemented experimentally~\cite{Ertmer,Denschlag}.
For small particle interactions, the self-consistent excited states (such as a
p-wave or d-wave condensate with at least one stationary dark soliton) approach
the non-interacting single-particle excitations of the harmonic trap. In this
regime, it might be possible to transfer most of the condensate into a soliton
state by applying an external field resonant with the energy difference between
the ground and excited self-consistent states.

We have attempted to excite the $d_{xy}$ dark soliton state for a small
condensate containing $N_0=1024$ atoms in a completely anisotropic trap, shown
in Fig.~\ref{pdwave}(b). The field excitation is modeled by a large-amplitude
time-dependent spatial perturbation in the GP equation~(\ref{GP}) given by
$V_{\rm add}({\bf{r}},t) = A(t)xy\cos(\omega_p t)$, where the amplitude
$A(t)$ is $25\%$ of a trap energy $\hbar\omega_x$ and includes a smooth turn
on and turn off as a function of time. The probe frequency is set at
$\omega_p=2.06$, which is the separation in chemical potential of the ground
and first $d_{xy}$ state (cf.\ Table~\ref{dtable}).

\begin{figure}
\psfig{figure=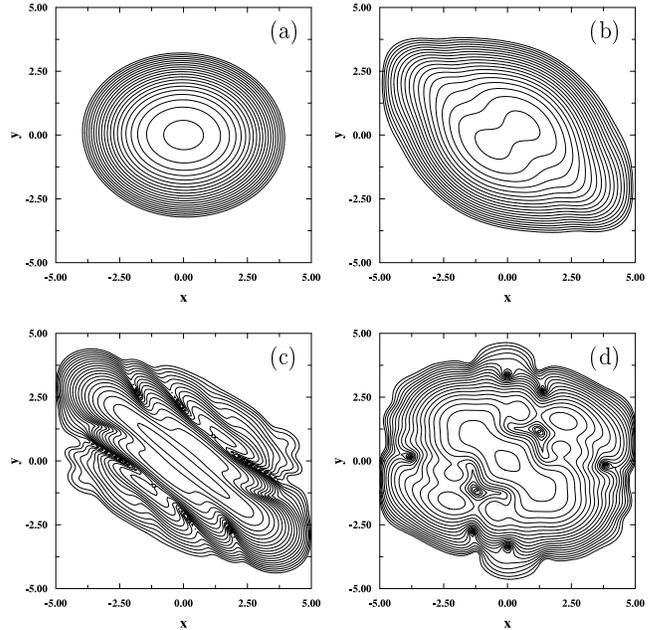,width=\columnwidth,angle=0}
\caption{The probability density in the $z = 0$ plane for the time evolution of
the ground state with $N = 1024$ under a time-varying spatial perturbation is
shown as a contour map at (a) $t = T_p\approx 0.5T$, (b) $t = 4T_p$, (c)
$t = 7T_p$, and (d) $t = 10T_p$. Radial distances are in scaled trap units
$d_x$, and trap parameters are $\omega_x=(2\pi)177$~rad/s, $\alpha=\sqrt{2}$,
and $\beta=2$. In (d), density minima at the top and bottom correspond to
vortices, those at left and right to antivortices.}
\label{vortices}
\end{figure}

The time-dependent probability density in the $z = 0$ plane is shown in
Fig.~\ref{vortices} with snapshots at $t=T_p$, $4T_p$, $7T_p$, and $10T_p$,
where $T_p=2\pi/\omega_p\approx 0.5T$. At $t = T_p$, the wavefunction overlap
with the ground state is $0.98$ and with the $d_{xy}$ state is $0.01$. Until
$t = 4T_p$, the wavefunction overlap with the ground state decreases
monotonically to $0.57$, while the overlap with the $d_{xy}$ state increases
to a maximum of $0.25$. Thereafter, as illustrated at times $t = 7T_p$ and
$t = 10T_p$, the wavefunction overlaps with both the ground and this particular
dark soliton state decrease. While the perturbing potential is not able to
yield a $d_{xy}$ state, the last image in Fig.~\ref{vortices} at $t = 10T_p$
shows the formation of a number of depressions in the probability density at
the very edge of the condensate, corresponding to multiple vortices with
both senses of circulation.

The inability of a field excitation to produce dark solitons in not
surprising, in light of the snake instabilities discussed in
Sec.~\ref{sec:snake}. Dark solitons are unstable against the formation of
vortices under ideal conditions, and are increasingly unstable in the
presence of an external perturbation.  Nevertheless, applying a very large
amplitude perturbation may be a viable alternative approach for the formation
of vortices in systems with much larger condensate densities; this possibility
remains for future work.

\section{conclusions}
\label{sec:conclusions}

In summary, self-consistent ground and excited states of a condensate in an
anisotropic harmonic trap are calculated by direct solution of the
time-dependent GP equation in three dimensions. The energy of a dark soliton
at the center of the trap, relative to the true ground state, is found to be
independent of both the number of atoms and the trap geometry in the TF limit,
with a value equal to the soliton oscillation frequency. In the weakly
interacting limit, however, the energy of the anomalous mode does not equal 
the energy of the dark soliton. In both limits, the low-lying Bogoliubov
excitation spectrum of p-wave states is found to contain modes with complex
frequencies, which may be removed by strong trap confinement in the soliton
plane or by decreasing the condensate density. These complex modes are shown
to give rise to the snake instability of the solitons observed in real time
propagation of the GP equation. In extended trap geometries, the solitons are
found to decay into vortex lines and rings at long times, imposing constraints
on the ability to generate and observe these intriguing excitations in current
experimental geometries.

\begin{acknowledgments}

The authors are grateful to G.~M.~Bruun, P.~Colarusso, P.~S.~Julienne,
W.~P.~Reinhardt, and J.~Simsarian for stimulating discussions, and to
P.~Ketcham for his assistance in generating the color figures. This work was
supported in part by the National Science Foundation (MSP), and by the U.S.\
Office of Naval Research (DLF and CWC). Part of the computational work was
carried out at the National Energy Research Supercomputer Center at Lawrence
Berkeley National Laboratory. Work (LAC) performed under the auspices of the
U.S.\ Department of Energy through the Los Alamos National Laboratory.

\end{acknowledgments}

\begin{table}
\caption{The chemical potential $\mu$ and free energy per particle $E$ of the
ground state (subscript $0$) and p-wave dark solitons oriented along $x$,
$y$, and $z$ are given as a function of the number of atoms in the condensate
$N_0$ in units of $\hbar\omega_x$.}
\begin{tabular}{ccccccccc}
$N_0$ & $\mu_0$ & $E_0$ & $\mu_x$ & $E_x$ & $\mu_y$ & $E_y$ & $\mu_z$ & $E_z$ \\
\hline \noalign{\vskip0.1cm}
$2^{10}$ & 3.57 & 2.99 & 4.40 & 3.87 & 4.79 & 4.27 & 5.36 & 4.85 \\
$2^{12}$ & 5.42 & 4.20 & 6.20 & 5.02 & 6.56 & 5.40 & 7.09 & 5.95 \\
$2^{14}$ & 8.90 & 6.59 & 9.64 & 7.36  & 9.98 & 7.71 & 10.47 & 8.23 \\
$2^{16}$ & 15.13 & 10.96 & 15.85 & 11.70 & 16.17 & 12.03 & 16.63 & 12.52 \\
$2^{18}$ & 26.10 & 18.75 & 26.80 & 19.47 & 27.12 & 19.79 & 27.57 & 20.25 \\ 
$2^{20}$ & 45.28 & 32.41 & 45.94 & 33.10 & 46.28 & 33.43 & 46.72 & 33.88 \\
\end{tabular}
\label{ptable}
\end{table}

\begin{table}
\caption{The chemical potential $\mu$ and free energy per particle $E$ of the
ground state (subscript $0$) and d-wave dark solitons with nodes along
$(x,y)$, $(x,z)$, and $(y,z)$ are given as a function of the number of atoms
in the condensate $N_0$ in units of $\hbar\omega_x$.}
\begin{tabular}{ccccccccc}
$N_0$ & $\mu_0$ & $E_0$ & $\mu_{x,y}$ & $E_{x,y}$ & $\mu_{x,z}$ & $E_{x,z}$
& $\mu_{y,z}$ & $E_{y,z}$ \\
\hline \noalign{\vskip0.1cm}
$2^{10}$ & 3.57 & 2.99 & 5.63 & 5.16 & 6.20 & 5.74 & 6.60 & 6.15 \\
$2^{12}$ & 5.42 & 4.20 & 7.33 & 6.22 & 7.87 & 6.77 & 8.23 & 7.16 \\
$2^{14}$ & 8.90 & 6.59 & 10.72 & 8.48 & 11.21 & 9.00 & 11.55 & 9.35 \\
$2^{16}$ & 15.13 & 10.96 & 16.89 & 12.77 & 17.35 & 13.26 & 17.67 & 13.59 \\
$2^{18}$ & 26.10 & 18.75 & 27.83 & 20.51 & 28.27 & 20.97 & 28.59 & 21.29 \\
$2^{20}$ & 45.28 & 32.41 & 46.94 & 34.12 & 47.38 & 34.57 & 47.73 & 34.90 \\
\end{tabular}
\label{dtable}
\end{table}

\end{document}